\documentclass{article}
\usepackage{amsmath,graphicx,hyperref}
\usepackage{booktabs}
\usepackage{multirow}
\usepackage[preprint]{spconf}
\usepackage[table,xcdraw]{xcolor}

\title{Speaker Anonymisation for Speech-based Suicide Risk Detection}

\name{
\shortstack{
Ziyun Cui$^{1,2}$, Sike Jia$^{2}$, Yang Lin$^{4}$, Yinan Duan$^{3}$, Diyang Qu$^{3}$,\\ 
\textit{Runsen Chen}$^{3}$, \textit{Chao Zhang}$^{1,2}$, \textit{Chang Lei}$^{3,\dagger}$, \textit{Wen Wu}$^{1,\dagger}$
\thanks{$^\dagger$ Corresponding authors. \\This work was supported by the National Natural Science Foundation of China (Grant No. 62501336).}
}
}
\address{$^{1}$Shanghai Artificial Intelligence Laboratory, Shanghai, China \\
$^{2}$Department of Electronic Engineering, Tsinghua University, Beijing, China \\
$^{3}$Vanke School of Public Health, Tsinghua University, Beijing, China \\
$^{4}$Weiyang College, Tsinghua University, Beijing, China \\
\small{\texttt{leic22@mails.tsinghua.edu.cn, cz277@tsinghua.edu.cn, wuwen@pjlab.org.cn}} 
  }

\begin{document}

\copyrightnotice{$\copyright$~IEEE 2026}

\ninept

\maketitle

\begin{abstract}
Adolescent suicide is a critical global health issue, and speech provides a cost-effective modality for automatic suicide risk detection. 
Given the vulnerable population, protecting speaker identity is particularly important, as speech itself can reveal personally identifiable information if the data is leaked or maliciously exploited.
This work presents the first systematic study of speaker anonymisation for speech-based suicide risk detection. 
A broad range of anonymisation methods are investigated, including techniques based on traditional signal processing, neural voice conversion, and speech synthesis. 
A comprehensive evaluation framework is built to assess the trade-off between protecting speaker identity and preserving information essential for suicide risk detection.
Results show that combining anonymisation methods that retain complementary information yields detection performance comparable to that of original speech, while achieving protection of speaker identity for vulnerable populations.

\end{abstract}
\begin{keywords}
speaker anonymisation, suicide risk detection, adolescent speech
\end{keywords}

\section{Introduction}

Suicide is a critical global health challenge and one of the leading causes of death among adolescents~\cite{world2025suicide, shain2016suicide}. Early detection of suicide risk is essential for effective prevention and intervention of potential suicide attempts.  Existing approaches still face limitations: clinical interviews require well-trained professionals, while self-report questionnaires are vulnerable to bias, discrimination, and even intentional concealment~\cite{ganzini2013trust}. Speech serves as a cost-effective, remote, and non-invasive candidate for automatic suicide risk detection~\cite{wu20251st}. Speech naturally encodes both semantic and paralinguistic information. As an individual becomes pre-suicide, their speech exhibits measurable alterations in acoustic properties (\textit{e.g.}, jitter, shimmer, fundamental frequency)~\cite{cummins2015review}, increased disfluency (\textit{e.g.}, hesitations and speech errors)~\cite{stasak2021read}, and distinct lexical patterns~\cite{belouali2021acoustic}. Previous studies have shown that suicide risk is detectable from spontaneous speech~\cite{wu20251st,cui2024spontaneous}.

Despite its great potential, it is critical to emphasise the privacy protection of participants' speech data, especially given the dual vulnerabilities of age and psychological distress in the relevant population. 

\begin{figure}[t]
    \centering
    \includegraphics[width=\linewidth]{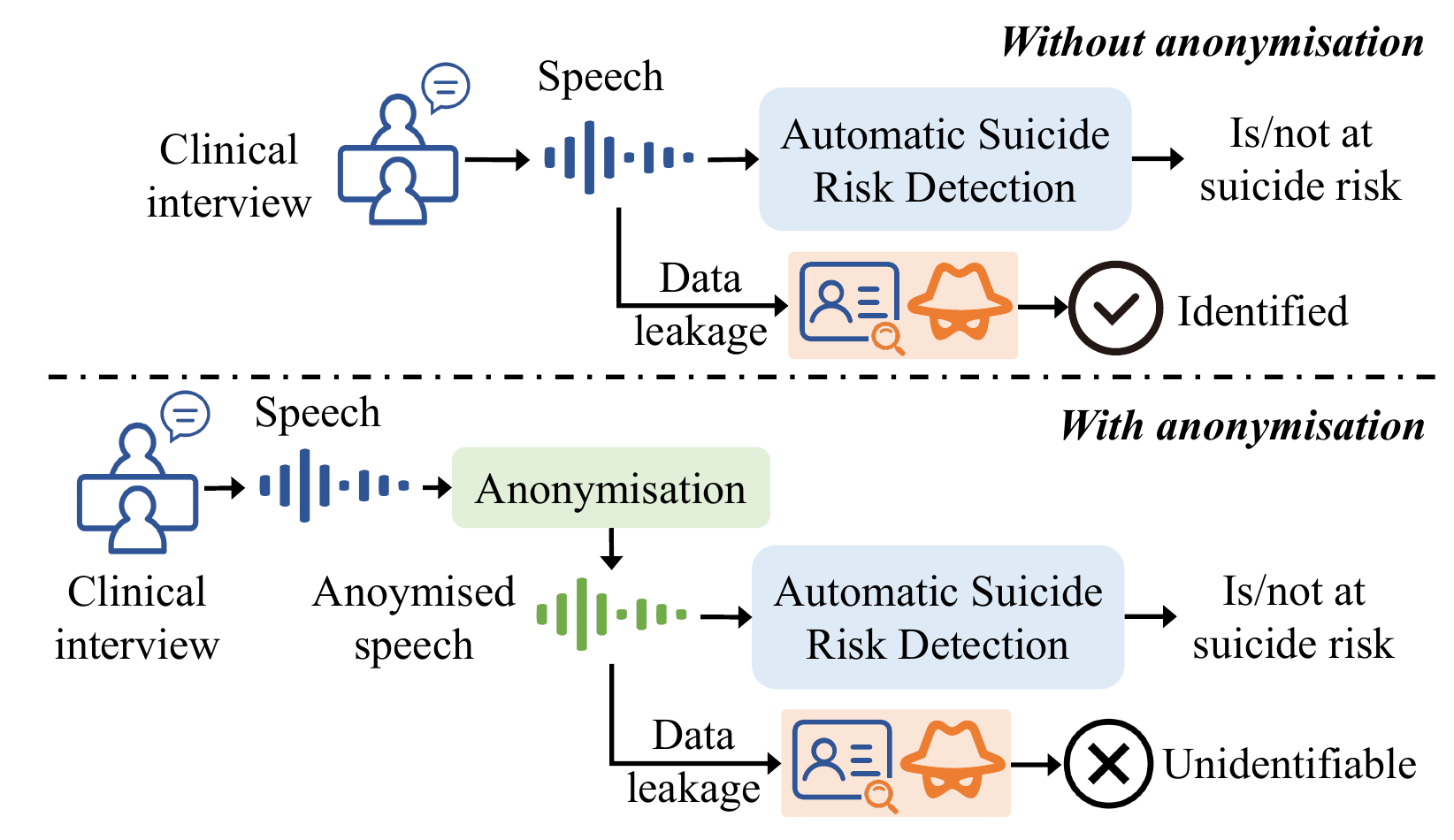}
    \caption{Importance of anonymisation for protecting speaker identity.}
    \label{fig: pipeline}
    \vspace{-2ex}
\end{figure}

As illustrated in Fig.~\ref{fig: pipeline}, 
directly incorporating the original speech into the automatic detection system could pose serious privacy risks, as the speaker’s identity can be easily revealed if the data is leaked or maliciously exploited. 
By contrast, applying proper anonymisation before exposing speech to the system makes it much harder to trace the processed audio back to individuals, even in the event of a data breach.
In fact, privacy protection has long been a central concern in the healthcare research, where anonymisation has been extensively studied for textual records~\cite{li2017anonymizing}, medical imaging~\cite{hellmann2024ganonymization}, and physiological signals such as ECG~\cite{nolin2023privecg} and EEG~\cite{matovu2016your}, where techniques range from simple de-identification~\cite{li2017anonymizing, matovu2016your} to advanced generative models~\cite{hellmann2024ganonymization, nolin2023privecg}. Speech data poses unique challenges: it simultaneously encodes linguistic, acoustic, and emotional information, many of which are relevant for clinical analysis but also carry identifiable speaker characteristics~\cite{tayebi2024addressing}, making it essential to carefully evaluate anonymisation techniques on both effectiveness in suppressing speaker identity and the ability to preserve features relevant to downstream suicide risk detection.

This work investigates speaker anonymisation for speech-based suicide risk detection. The aim is to identify anonymisation pipelines that strike a balance between sufficient privacy protection and performance in suicide risk detection.
We systematically evaluate a wide range of anonymisation methods, including traditional signal processing techniques, neural voice conversion  approaches, and speech synthesis approaches.
A comprehensive evaluation framework is built to assess the effect of anonymisation approaches in terms of speech quality, speaker discrimination, deviation in fundamental frequency, as well as preservation of semantic and emotional content.
Experiments show that combining complementary anonymisation methods boosts downstream task performance, achieving detection accuracy comparable to that of original speech while still maintaining robust anonymisation.

\section{Methods}
\subsection{Speech-based Suicide Risk Detection}

\begin{figure}[h]
    \centering
    \includegraphics[width=\linewidth]{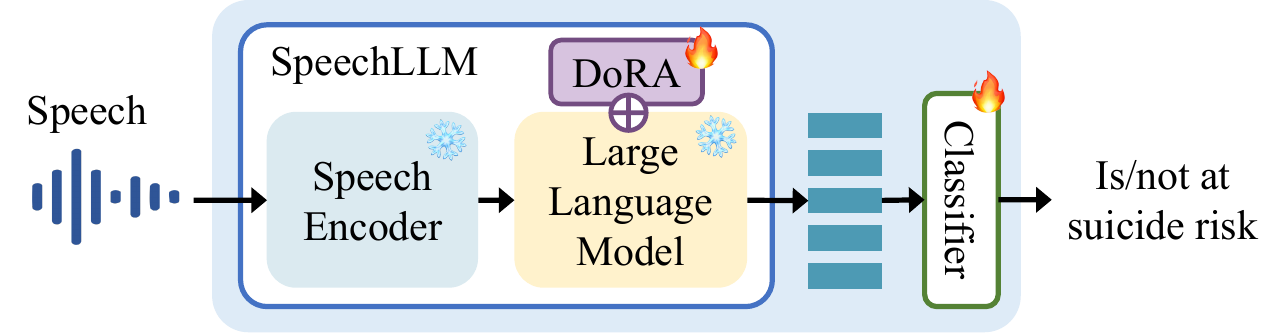}
    \vspace{-4ex}
    \caption{Structure used for speech-based suicide risk detection.}
    \label{fig: classify-pipeline}
\end{figure}

This section describes the structure used in this study for speech-based suicide risk detection. 
The task is formulated as binary classification. As shown in Fig.~\ref{fig: classify-pipeline}, a speech LLM (Qwen2.5-Omni-7B~\cite{xu2025qwen2}\footnote{\url{https://huggingface.co/Qwen/Qwen2.5-Omni-7B}}), which consists of a speech encoder and a large language model (LLM), is employed as the backbone for processing speech recordings and generating embeddings for downstream suicide risk detection.
Weight-decomposed low-rank adaptation (DoRA) is adopted for parameter-efficient fine-tuning of the speech LLM, with a rank of 32 and alpha of 64. 

We perform speech-based suicide risk detection using original and anonymised speech respectively to examine whether using anonymised speech leads to remarkable performance degradation. A two-sided t-test between the results obtained with original and anonymised speech is conducted to assess statistical significance.

\subsection{Speaker Anonymisation Approaches}
This section describes the anonymisation approaches studied in this paper, including traditional signal processing methods, voice conversion (VC) based on neural vocoder, and direct synthesis based on transcribed text.

\subsubsection{Traditional Signal Processing Methods}

Traditional anonymisation methods involves direct transformation of acoustic attributes without altering the linguistic content, which are computationally lightweight and interpretable. Two approaches are studied in this paper:

\textbf{Pitch modification}  alters fundamental frequency (F0) of speech without significantly distorting prosodic and speaking style. 
The magnitude of the shift, measured in semitones, affects the trade-off among anonymisation effectiveness, naturalness, and utility.

\textbf{McAdams}~\cite{patino2021speaker} leverages Linear Predictive Coding (LPC) to alter vocal tract characteristics by shifting the formant locations. 
The LPC order is a critical parameter that governs the trade-off between the fidelity of the speech signal and the degree of anonymisation. 

\subsubsection{Content-Speaker Disentanglement with Neural Vocoder}

This section investigates approaches that leverage the paradigm of content-speaker disentanglement. They first extract content representations, then manipulate or replace speaker-specific embeddings, and finally synthesise speech from the modified representations by a neural vocoder. 

\textbf{SSL-SAS}~\cite{miao2022analyzing} 
employs a HuBERT-based content encoder to extract frame-level content features, ECAPA-TDNN for speaker encoder, an F0 contour extraction, before a neural vocoder HiFi-GAN synthesising the final anonymised speech. The target speaker embedding is a pseudo-speaker vector constructed by averaging 100 similar speaker embeddings from an external pool derived from the LibriTTS dataset.

\textbf{FreeVC}~\cite{li2023freevc} implements text-free, one-shot voice conversion using WavLM features as rich speech representations and an information bottleneck mechanism to extract clean content embeddings that minimize speaker identity leakage. 

\textbf{SeedVC}~\cite{liu2024zero} is a Diffusion-based voice conversion framework, which leverages Whisper-extracted content features to preserve linguistic information and CAM++ speaker features to capture target speaker characteristics, which serve as U-Net conditional inputs to guide the diffusion generation process. The diffusion process iteratively refines the output through multiple denoising steps. 

\textbf{RVC}\footnote{\url{https://github.com/RVC-Project/Retrieval-based-Voice-Conversion-WebUI}} combines retrieval mechanisms with neural generation for speaker transformation. RVC retrieves segments or acoustic embeddings from a target-speaker database to guide conversion and synthesises output via a neural vocoder, thereby enhancing audio fidelity and naturalness through the reuse of real acoustic patterns. 

\subsubsection{Speech Synthesis from Transcribed Text}
For comparison, we investigate a cascaded approach that first conducts automatic speech recognition (ASR) to transcribe the original speech to text, and then synthesises new speech from the transcribed text using text-to-speech (TTS) technology. This method fundamentally disrupts the original speaker's identity by completely discarding all acoustic, prosodic, and emotional characteristics present in the source speech, while attempts to preserve the semantic content. We employ Paraformer~\cite{gao2022paraformer, gao2023funasr} for ASR, and evaluate two popular TTS models: \textbf{SparkTTS}~\cite{wang2025spark}\footnote{\url{https://huggingface.co/SparkAudio/Spark-TTS-0.5B}} and \textbf{CosyVoice 2.0}~\cite{du2024cosyvoice}\footnote{\url{https://huggingface.co/FunAudioLLM/CosyVoice2-0.5B}}.

\subsection{Anonymisation Performance Evaluation Framework}
\label{sec:metric}

We build a multi-dimensional evaluation framework to comprehensively assess the effectiveness of different anonymisation methods, spanning over the following five critical aspects: speech quality, speaker discrimination, deviation in fundamental frequency, preservation of semantic content, and preservation of emotional content.

\textbf{Speech Quality.} Two metrics are employed to measure the quality of anonymised speech: signal-to-noise ratio (\textbf{SNR}) and MOS score. SNR measures cleanliness of the speech while MOS measures the overall perceptual quality of the speech. The MOS score is automatically computed by the UTMOS model~\cite{saeki2022utmos}.

\textbf{Speaker Discrimination.} The discrimination between speakers before and after anonymisation is a key aspect in evaluating the effectiveness of anonymisation.
Equal Error Rate (\textbf{EER}) is used to assess speaker verifiability after anonymisation, which  
reflects the traceability of anonymised speech and its resistance to linking attacks. 

\textbf{Deviation in Fundamental Frequency}. We perform F0 contour variation analysis to quantify the pitch change after anonymisation. F0 contours are extracted from both original and anonymised speech. Two complementary metrics are computed between the contours: (i) \textbf{L1} distance, which captures the average magnitude of pitch shifts; (ii)  Pearson correlation coefficient (\textbf{PCC}), which quantifies the degree to which the pitch-contour shape is preserved. 

\textbf{Preservation of Semantic Content.} The semantic consistency is evaluated by character error rate (\textbf{CER}) between the original and anonymised speech, both transcribed by Paraformer ASR system.

\textbf{Preservation of Emotional Content.} Since emotional information might contain important cue for suicide risk assessment, the preservation of it after anonymisation is also evaluated. 
Emotion embeddings are extracted using
Emotion2Vec~\cite{ma2024emotion2vec}\footnote{\url{https://huggingface.co/emotion2vec/emotion2vec_plus_seed}} from both original and anonymised speech samples. Emotion consistency is then measured by the cosine similarity. 

\begin{table}[t]
\centering
\caption{Comparison of different speaker anonymisation methods under different configurations. Configuration used for later experiments shown in bold.}
\label{tab: anonymise audio}
\vspace{1ex}
\resizebox{\linewidth}{!}{
\setlength{\tabcolsep}{3pt} 
\begin{tabular}{@{}cccccccc@{}}
\toprule
                                & \textbf{SNR}$\uparrow$ & \textbf{MOS}$\uparrow$ & \textbf{L1$_\text{F0}$}$\downarrow$ & \textbf{PCC$_\text{F0}$}$\uparrow$ & \textbf{CER}$\downarrow$ & \textbf{Emo}$\uparrow$ & \textbf{EER}$\uparrow$ \\ \midrule
\textbf{SparkTTS}                       & 35.96                  & 3.11                   & 9.260                               & 0.009                              & 0.332                    & 0.577                  & 0.498                  \\
\textbf{CosyVoice}                       & 61.89                  & 3.01                   & 7.467                               & -0.002                             & 0.024                    & 0.257                  & 0.497                  \\ \midrule
Pitch$_\text{step2}$            & 15.05                  & 1.31                   & 3.001                               & 0.563                              & 0.343                    & 0.835                  & 0.326                  \\
\textbf{Pitch$_\text{step4}$}   & 15.00                  & 1.32                   & 4.815                               & 0.526                              & 0.424                    & 0.836                  & 0.512                  \\
Pitch$_\text{step6}$            & 14.97                  & 1.31                   & 6.495                               & 0.480                              & 0.540                    & 0.837                  & 0.508                  \\ \midrule
McAdams$_\text{lpc15}$          & 14.92                  & 1.28                   & 1.761                               & 0.587                              & 0.335                    & 0.858                  & 0.248                  \\
\textbf{McAdams$_\text{lpc20}$} & 15.78                  & 1.28                   & 2.837                               & 0.510                              & 0.457                    & 0.851                  & 0.295                  \\
McAdams$_\text{lpc25}$          & 16.11                  & 1.28                   & 4.039                               & 0.421                              & 0.603                    & 0.848                  & 0.331                  \\ \midrule
\textbf{SSL-SAS$_\text{F0}$}    & 47.18                  & 2.25                   & 2.328                               & 0.377                              & 0.351                    & 0.805                  & 0.355                  \\
SSL-SAS$_\text{no\_F0}$         & 34.49                  & 1.33                   & 13.08                               & 0.122                              & 0.388                    & 0.661                  & 0.475                  \\ \midrule
\textbf{FreeVC$_\text{child}$}  & 24.40                  & 2.31                   & 4.984                               & 0.372                              & 0.367                    & 0.868                  & 0.445                  \\
FreeVC$_\text{adult}$           & 28.88                  & 2.61                   & 9.642                               & 0.421                              & 0.362                    & 0.870                  & 0.453                  \\ \midrule
\textbf{SeedVC$_\text{d25}$}    & 37.58                  & 2.09                   & 7.215                               & 0.299                              & 0.209                    & 0.820                  & 0.492                  \\
SeedVC$_\text{d40}$             & 36.35                  & 2.05                   & 7.225                               & 0.302                              &    0.207                    & 0.870                  & 0.499                  \\ 
\midrule
\textbf{RVC$_\text{s1}$}        & 39.28                  & 2.31                   & 1.672                               & 0.490                              & 0.398                    & 0.794                  & 0.510                  \\
RVC$_\text{s2}$                 & 54.18                  & 2.47                   & 2.730                               & 0.443                              & 0.362                    & 0.619                  & 0.501                  \\ \midrule
Original                        & 15.24                  & 1.49                   & -                                   & -                                  & -                        & -                      & 0.185                  \\ \bottomrule
\end{tabular}
}
\end{table}

\section{Experimental Setup}

\subsection{Dataset}

The dataset used in this study consists of voice recordings collected from 1,223 Chinese adolescents aged 10-18~\cite{wu20251st, cui2024spontaneous}. All recordings were conducted in Mandarin Chinese. Recordings are all paired with standardised suicide risk assessments administered via the Mini International Neuropsychiatric Interview for Children and Adolescents (MINI-KID) suicidality module~\cite{sheehan2010reliability}, a widely validated diagnostic interview tool. 53.4\% participants were identified as in suicide risk. The data collection procedures have been approved by the Medical Ethics Committee of the Tsinghua University Technology Ethics Committee. Following previous study~\cite{cui2024spontaneous}, the recordings corresponding to the self-description task is used.

Given the sensitive nature of mental health data and the vulnerable population involved, protecting adolescent privacy through effective speech anonymisation is important.  The selected self-description task contains rich acoustic and semantic information relevant to suicide risk detection, making it suitable for evaluating how well different anonymisation methods preserve both types of information. 

\subsection{Implementation Details}
The dataset was split into train/dev/test sets with an 8:1:1 ratio. The model achieving the best validation performance is selected for test evaluation. All experiments were conducted with three random seeds and the average accuracy is reported.

Below describes different configurations tested for each speaker anonymisation method.
For \textbf{pitch} modification, three different shift magnitudes (2, 4 and 6 semitones) were tested.
Male voices were raised by the number of semitones while female voices were lowered by the same number. 
For \textbf{McAdams}, LPC with orders of 15, 20, and 25 were compared.
For \textbf{SSL-SAS}, we tested both with and without F0 contour preservation.
For \textbf{FreeVC}, we tested two target voice styles: a child voice and a male adult voice.
For \textbf{SeedVC}, we tested the diffusion step of 10 and 25. 
For \textbf{RVC}, two different teenager voices from the official release were tested, denoted as subscript \textit{s1} and \textit{s2} in Table~\ref{tab: anonymise audio}.

\begin{figure}[t]
    \centering
    \includegraphics[width=\linewidth]{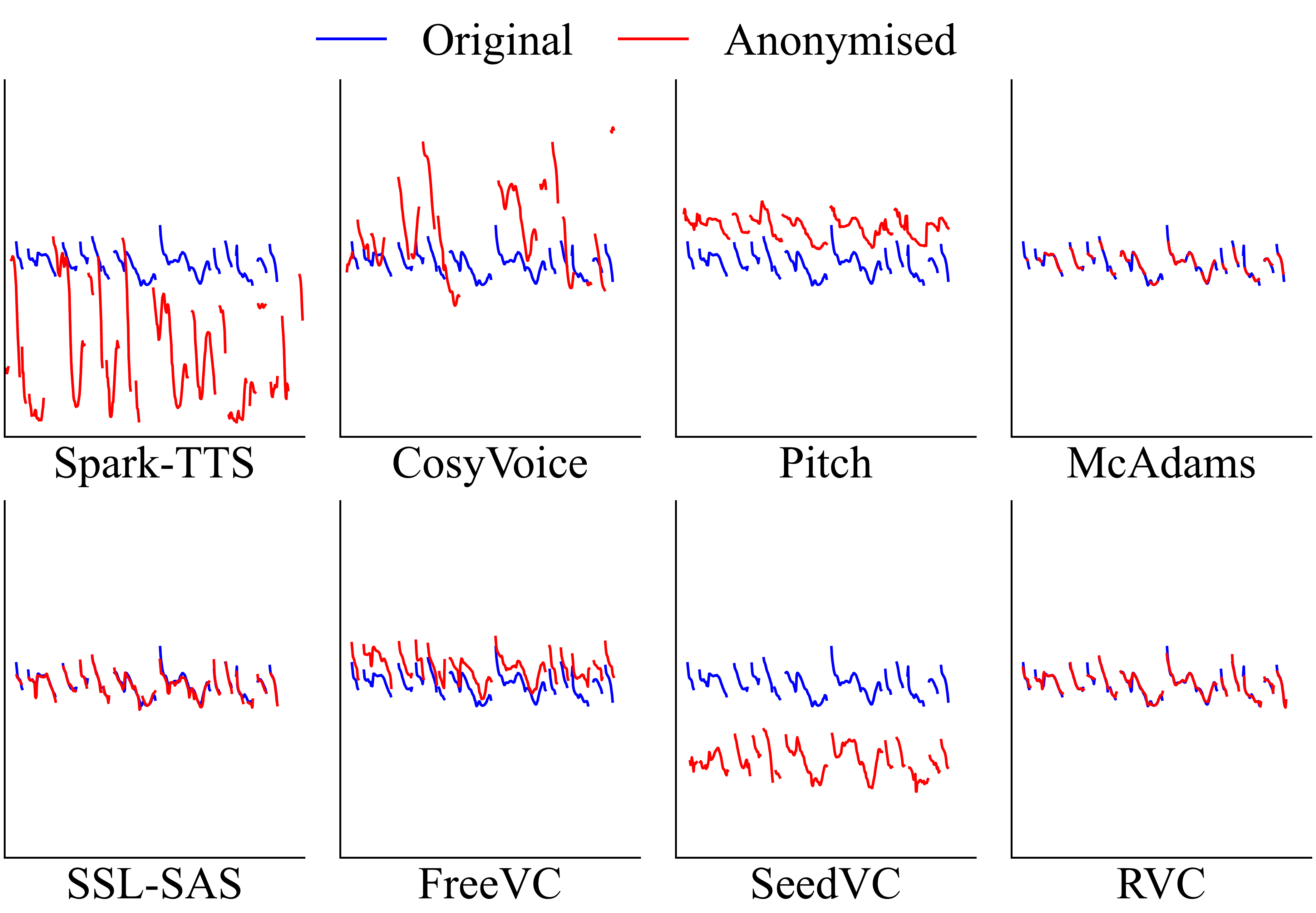}
    \vspace{-4ex}
    \caption{Visualisation of pitch change in semitones relative to A4 (440 Hz).}
    \label{fig:pitch}
\end{figure}

\section{Results and Discussion}

\subsection{Anonymisation Effectiveness and Perceptual Quality}

Table~\ref{tab: anonymise audio} compares various anonymisation methods across different configurations under the evaluation framework introduced in Section~\ref{sec:metric}. 
We also provide a visualisation of changes in F0 contours induced by different anonymisation methods in Fig.~\ref{fig:pitch}.

Firstly, methods that directly synthesise speech from transcribed text (\textit{i.e.,} SparkTTS and CosyVoice) achieve superior performance in preserving semantic content, as indicated by lower CER, and produce better perceptual quality, as reflected by higher UTMOS.
However, they completely discard acoustic and emotional characteristics, yielding low PCC$_\text{F0}$ and emotion similarity, as expected. This is evident in Fig.~\ref{fig:pitch}, where the F0 contours of anonymised speech are extremely different from the originals.

For traditional signal processing, both pitch shifting and McAdams yield relatively low signal quality, indicated by lower SNR. It is worth noting that pitch shifting exhibits large L1 deviations but high PCC in F0 contours. This seemingly counter-intuitive pattern is in fact expected, as it indicates that only the absolute F0 values are shifted while contour shapes are preserved, as shown in Fig.~\ref{fig:pitch}.
It can be also observed that increase in shifting step leads to higher L1 deviations and lower PCC. 
The \textit{step=4} configuration provides overall the best anonymisation trade-off. 
A similar pattern is observed for McAdams when LPC step increases. While McAdams preserves F0 contours and emotional content well, it performs poorly in speaker discrimination, as indicated by the lowest EER among the anonymisation approaches.

All four VC-based methods improve the speech quality compared to original speech, with higher SNR and UTMOS. Among four methods, SeedVC yields the lowest CER while SSL-SAS gives the lowest EER. It can be observed that without F0 preservation, SSL-SAS$_\text{no\_F0}$ results in much larger L1$_\text{F0}$ deviation, much lower PCC$_\text{F0}$, as well as consistently poorer perceptual quality, intelligibility, and emotion similarity. 
Among all tested speaker anonymisation approach, 
FreeVC provides the best emotional consistency while RVC achieves the best speaker discrimination performance as well as the smallest F0 deviation.

Based on the overall trade-off between anonymisation effectiveness and speech quality, one configuration is selected for each anonymisation approach, bolded in Table~\ref{tab: anonymise audio} and used in the following experiments.

\subsection{Suicide Risk Detection Using Anonymised Speech}

\begin{figure}[t]
    \centering
    \includegraphics[width=0.95\linewidth]{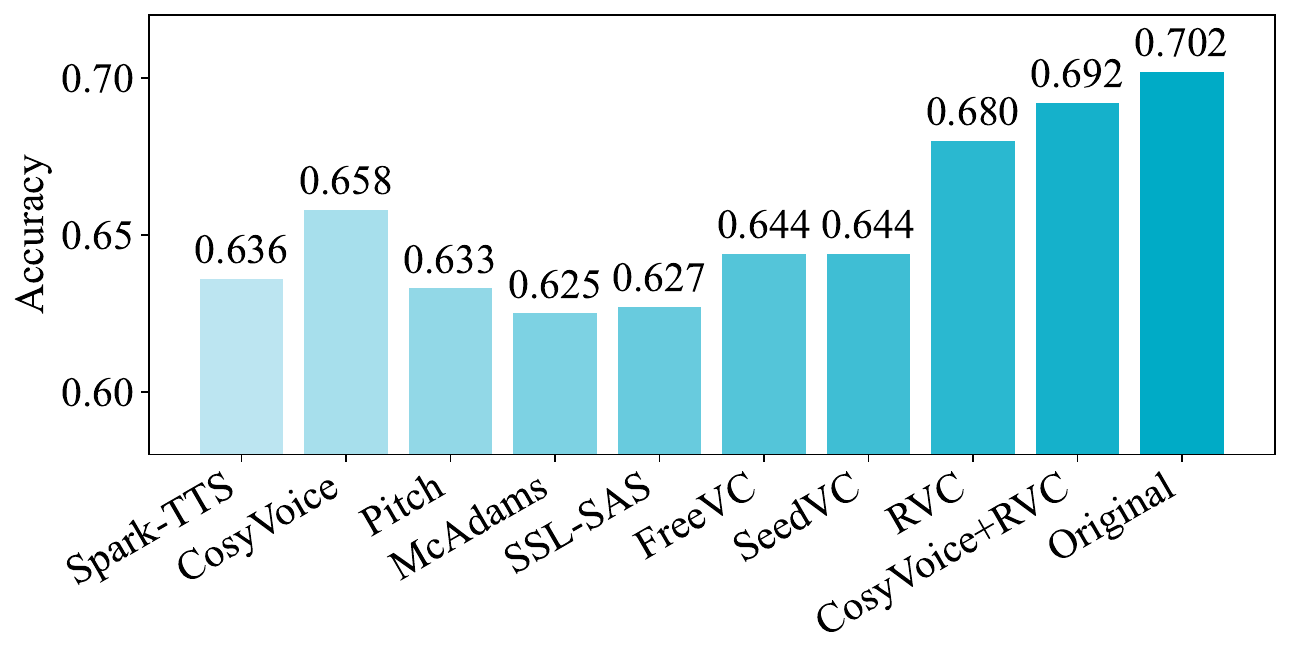}
    \vspace{-2ex}
    \caption{Accuracy of suicide risk detection using the original and anonymised speech. Average of three runs reported.}
    \label{fig:acc}
\end{figure}

Fig.~\ref{fig:acc} shows the suicide risk detection results using speech anonymised by different approaches.
The original speech got an average accuracy of 0.702, serving as the upper-bound. 
Among all anonymisation methods, RVC produced the strongest single-model performance with an accuracy of 0.680, only 2\% drop compared to original audio (with p=0.349), while maintaining high anonymisation strength (with EER of 0.510). CosyVoice achieved the second-best result (accuracy of 0.658, 5\% decay, with p=0.136), possibly benefiting from the best preservation of semantic content while suffering from the loss of acoustic and prosodic cues. All other approaches show significant performance decay.

We further ensemble the CosyVoice and RVC models by averaging their predicted probabilities. This ensemble benefits from the complementary strengths of the two approaches where CosyVoice retains most semantic contents and RVC maintains most F0 contour. The combined system achieves an accuracy of 0.692. This result surpasses all individual systems, reduces the gap to the original speech to only ~1\% (with p=0.677), while still ensuring good anonymisation. These findings suggest that combining complementary anonymisation strategies enhances speech utility while preserving privacy.

\subsection{The Impact of Speech Enhancement}

\begin{table}[t]
\centering
\caption{Performance of anonymisation approaches on enhanced. Results improved over non-enhanced ones (Table~\ref{tab: anonymise audio}) marked in bold.}
\vspace{1ex}
\label{tab: enhance audio}
\resizebox{\linewidth}{!}{
\setlength{\tabcolsep}{3pt} 
\begin{tabular}{@{}cccccccc@{}}
\toprule
                     & \textbf{SNR}$\uparrow$ & \textbf{MOS}$\uparrow$ & \textbf{L1$_\text{F0}$}$\downarrow$ & \textbf{L1$_\text{PCC}$}$\uparrow$ & \textbf{CER}$\downarrow$ & \textbf{Emo}$\uparrow$ & \textbf{EER}$\uparrow$ \\ \midrule
SparkTTS            & \textbf{36.14}         & 2.96                   & 9.493                               & 0.006                              & \textbf{0.336}           & 0.474                  & \textbf{0.506}         \\
CosyVoice            & 61.79                  & \textbf{3.10}          & 7.412                               & \textbf{0.002}                     & 0.031                    & 0.213                  & \textbf{0.505}         \\
Pitch$_\text{step4}$   & \textbf{50.12}         & 1.26                   & \textbf{4.398}                      & \textbf{0.646}                     & \textbf{0.220}           & 0.833                  & \textbf{0.569}         \\
McAdams$_\text{lpc20}$ & \textbf{51.59}         & \textbf{1.29}          & \textbf{1.833}                      & 0.504                              & \textbf{0.224}           & 0.819                  & 0.263                  \\
SSL-SAS$_\text{F0}$    & \textbf{57.12}         & \textbf{2.53}          & \textbf{1.349}                      & \textbf{0.520}                     & \textbf{0.229}           & \textbf{0.833}         & \textbf{0.356}         \\
FreeVC$_\text{child}$  & \textbf{31.29}         & \textbf{2.50}          & \textbf{4.953}                      & \textbf{0.421}                     & \textbf{0.305}           & 0.833                  & 0.434                  \\
SeedVC$_\text{d25}$    & \textbf{41.65}         & \textbf{2.12}          & 7.259                               & \textbf{0.363}                     & \textbf{0.177}           & \textbf{0.848}         & 0.477                  \\
RVC$_\text{s1}$        & \textbf{47.66}         & \textbf{2.39}          & \textbf{0.737}                      & \textbf{0.685}                     & \textbf{0.202}           & 0.778                  & \textbf{0.511}         \\
Original             & \textbf{52.85}         & \textbf{2.23}          & -                                   & -                                  & -                        & -                      & 0.155                  \\ \bottomrule
\end{tabular}
}
\end{table}

\begin{table}[t]
\centering
\caption{Accuracy of suicide risk detection with enhanced speech. Results improved over non-enhanced ones (Fig.~\ref{fig:acc}) bolded.}
\label{tab: acc after enhance}
\vspace{1ex}
\resizebox{0.8\linewidth}{!}{
\begin{tabular}{cccc}
\toprule
 \textbf{SparkTTS}      & \textbf{CosyVoice}      & \textbf{Pitch}          & \textbf{McAdams}         \\
 0.625          & 0.644          & \textbf{0.642} & \textbf{0.633}  \\
\midrule
\midrule
 \textbf{SSL-SAS}        & \textbf{FreeVC}         & \textbf{SeedVC}         & \textbf{RVC}             \\
 0.622          & \textbf{0.664} & 0.636          & 0.664           \\
  \bottomrule
\end{tabular}
}
\vspace{-2ex}
\end{table}

It can be seen from Table~\ref{tab: anonymise audio} that the original speech has a relatively low SNR and UTMOS, which is a common phenomenon for raw clinical data. This section investigates whether applying speech enhancement benefits detection performance.
FRCRN~\cite{zhao2022frcrn} was applied to the original speech before performing anonymisation. 
Table~\ref{tab: enhance audio} shows the anonymisation performance on enhanced speech. 
For the two TTS-based methods, enhancement has little effect. For other methods, enhancement leads to improvement on most metrics.

However, enhancement does not lead to consistently improved suicide detection results, as shown in Table~\ref{tab: acc after enhance}.
Comparing with the detection performance of non-enhanced speech in Fig.~\ref{fig:acc}, except for the traditional methods, most of the rest approaches do not benefit from speech enhancement.
The cause can be twofold. 
The enhancement operation performs spectral masking, which, despite producing cleaner signals, can also lead to information loss from the original speech. In addition, 
modifications in spectral patterns may interfere with the encoders in VC models, which extracts relevant features.  This highlights the need to carefully consider compatibility between enhancement and anonymisation modules.

\section{Conclusion}

This work presents the first systematic investigation of speaker anonymisation methods for adolescent suicide risk detection, covering traditional signal processing, neural voice conversion, and speech synthesis approaches under a comprehensive evaluation framework. Results show that different anonymisation approaches excel at preserving different types of information, and combining complementary methods (\textit{i.e.}, RVC for acoustic features and CosyVoice for semantic content) achieves near-original performance, with only 1\% decay. These findings hightlight the potential of hybrid strategies for privacy-preserving clinical speech applications.

\bibliographystyle{IEEEbib}
\bibliography{refs}

\end{document}